\RequirePackage{ifpdf}
\ifpdf % We are running pdfTeX in pdf mode
\documentclass[pdftex]{sigma}
\else
\documentclass{sigma}
\fi

\begin{document}

\allowdisplaybreaks

\renewcommand{\PaperNumber}{022}

\FirstPageHeading

\ShortArticleName{Beyond the Gaussian}
\ArticleName{Beyond the Gaussian}

\Author{Kazuyuki FUJII}

\AuthorNameForHeading{K. Fujii}
\Address{Department of Mathematical Sciences,
  Yokohama City University,
  Yokohama, 236-0027 Japan}
\Email{\href{mailto:fujii@yokohama-cu.ac.jp}{fujii@yokohama-cu.ac.jp}}

\ArticleDates{Received January 12, 2011, in f\/inal form February 28, 2011;  Published online March 04, 2011}

\Abstract{In this paper we present a non-Gaussian integral based on a
  cubic polynomial, instead of a quadratic, and give a fundamental formula
  in terms of its discriminant.
  It gives a mathematical reinforcement to the recent result by Morozov
  and Shakirov.
  We also present some related results.
  This is simply one  modest step to go beyond the Gaussian
  but it already reveals many obstacles
  related with the big  challenge of going further beyond the Gaussian.}


\Classification{11D25; 11R29; 26B20; 81Q99}

\section{Introduction}

The {\it Gaussian} is an abbreviation of all subjects related to
the Gauss function $e^{-(px^{2}+qx+r)}$ like the Gaussian
beam, Gaussian process, Gaussian noise, etc.
It plays a fundamental role in mathematics, statistics,
physics and related disciplines.
It is generally conceived that any attempts to generalize the Gaussian results
would meet formidable dif\/f\/iculties.
Hoping to overcome this high wall of dif\/f\/iculties of going beyond the Gaussian
in the near future,   a f\/irst step was introduced in~\cite{Fu}. This paper is
its  polished  version.

In the paper \cite{MS} the following ``formula'' is reported:
\begin{gather}
\label{eq:cubic-integral}
\iint e^{-\left(ax^{3}+bx^{2}y+cxy^{2}+dy^{3}\right)}dxdy
= \frac{1}{\sqrt[6]{-D}},
\end{gather}
where $D$ is the discriminant
of the cubic equation
\[
ax^{3}+bx^{2}+cx+d=0,
\]
and it is given by
\begin{gather}
\label{eq:cubic-discriminant}
D=b^{2}c^{2}+18abcd-4ac^{3}-4b^{3}d-27a^{2}d^{2}.
\end{gather}
The formula (\ref{eq:cubic-integral}) is of course non-Gaussian.
However, if we consider it in the framework of the real category
then (\ref{eq:cubic-integral}) is not correct because the left hand side
diverges. In this paper we treat only the real category, and so $a$, $b$, $c$, $d$,
$x$, $y$ are real numbers.

Formally, by performing the change of variable $x=t\rho$, $y=\rho$
for (\ref{eq:cubic-integral}) we have
\begin{gather*}
\mbox{l.h.s. of (\ref{eq:cubic-integral})}
 =
\iint e^{-\rho^{3}\left(at^{3}+bt^{2}+ct+d\right)}|\rho|dtd\rho
=\int
\left\{\int e^{-\left(at^{3}+bt^{2}+ct+d\right)\rho^{3}}|\rho|d\rho\right\}dt
\\
\phantom{\mbox{l.h.s. of (\ref{eq:cubic-integral})}}{} =
\int |\sigma|e^{-\sigma^{3}}d\sigma
\int \frac{1}{\big|\sqrt[3]{(at^{3}+bt^{2}+ct+d)}\big|\sqrt[3]{(at^{3}+bt^{2}+ct+d)}}dt
\end{gather*}
by the change of variable $\sigma=\sqrt[3]{at^{3}+bt^{2}+ct+d} \rho$.

The divergence comes from
\begin{gather*}
  \int |\sigma|e^{-\sigma^{3}}d\sigma,
\end{gather*}
while the main part is
\begin{gather*}
  \int \frac{1}{\big|\sqrt[3]{(ax^{3}+bx^{2}+cx+d)}\big|\sqrt[3]{(ax^{3}+bx^{2}+cx+d)}}dx
\end{gather*}
under the change $t\rightarrow x$. As a kind of renormalization
the integral may be def\/ined like
\begin{gather*}
  \ddagger \iint_{{\mathbb R}^{2}}\! e^{-(ax^{3}+bx^{2}y+cxy^{2}+dy^{3})}dxdy\ \ddagger
  =
  \int_{{\mathbb R}}\!
  \frac{1}{\big|\sqrt[3]{(ax^{3}\!+bx^{2}\!+cx+d)}\big|\sqrt[3]{(ax^{3}\!+bx^{2}\!+cx+d)}}dx.
\end{gather*}
However, the right hand side lacks proper symmetry. If we set
\begin{gather*}
  F(a,b,c,d)=\iint_{D_{R}}e^{-\left(ax^{3}+bx^{2}y+cxy^{2}+dy^{3}\right)}dxdy,
\end{gather*}
where $D_{R}=[-R,R]\times [-R,R]$, then it is easy to see
\begin{gather*}
  F(-a,-b,-c,-d)=F(a,b,c,d).
\end{gather*}
Namely, $F$ is invariant under ${\mathbb Z}_{2}$-action.  This symmetry is
important and must be kept even in the renormalization process.
The right hand side in the ``def\/inition'' above is clearly not invariant.
Therefore, by modifying it slightly we reach the renormalized integral

\begin{definition}
\begin{gather}
\label{eq:renorm-definition}
\ddagger \iint_{{\mathbb R}^{2}}e^{-(ax^{3}+bx^{2}y+cxy^{2}+dy^{3})}dxdy\ \ddagger
=
\int_{{\mathbb R}} \frac{1}{\sqrt[3]{(ax^{3}+bx^{2}+cx+d)^{2}}}dx.
\end{gather}
\end{definition}

We believe that the def\/inition is not so bad (see the Section~\ref{section4}).

In the paper we calculate the right hand side of (\ref{eq:renorm-definition})
{\it directly}, which will give some interesting results and a new perspective.
The result gives a mathematical reinforcement to the result~\cite{MS}
by Morozov and Shakirov.

\section{Main result} \label{section2}

Before stating the result let us make some preparations.
The Gamma function $\Gamma(p)$ is def\/ined by
\begin{gather}
\label{eq:Gamma-function}
\Gamma(p)=\int_{0}^{\infty}e^{-x}x^{p-1}dx\qquad (p>0)
\end{gather}
and the Beta function $B(p,q)$ is
\begin{gather*}
%\label{eq:Beta-function}
B(p,q)=\int_{0}^{1}x^{p-1}(1-x)^{q-1}dx\qquad (p, q>0).
\end{gather*}
Note that the Beta function is rewritten as{\samepage
\begin{gather*}
  B(p,q)=\int_{0}^{\infty}\frac{x^{p-1}}{(1+x)^{p+q}}dx.
\end{gather*}
See \cite{WW} for more detail. Now we are in a position to
state the result.}

\noindent
{\bfseries Fundamental formula.}

(I) For $D < 0$
\begin{gather}
\label{eq:formula-I}
\int_{{\mathbb R}} \frac{1}{\sqrt[3]{(ax^{3}+bx^{2}+cx+d)^{2}}}dx
=
\frac{C_{-}}{\sqrt[6]{-D}},
\end{gather}
where
\begin{gather*}
  C_{-}=\sqrt[3]{2}B\left(\frac{1}{2},\frac{1}{6}\right).
\end{gather*}

(II)  For $D > 0$
\begin{gather}
\label{eq:formula-II}
\int_{{\mathbb R}} \frac{1}{\sqrt[3]{(ax^{3}+bx^{2}+cx+d)^{2}}}dx
=
\frac{C_{+}}{\sqrt[6]{D}},
\end{gather}
where
\begin{gather*}
  C_{+}=3B\left(\frac{1}{3},\frac{1}{3}\right).
\end{gather*}

(III) $C_{-}$ and $C_{+}$ are related by $C_{+}=\sqrt{3}C_{-}$
through the identity
\begin{gather}
\label{eq:formula-III}
\sqrt{3}B\left(\frac{1}{3},\frac{1}{3}\right)=\sqrt[3]{2}B\left(\frac{1}{2},\frac{1}{6}\right).
\end{gather}

Our result shows that the integral depends on the sign of $D$, and so
our question is as follows.

\medskip

\noindent
{\bfseries Problem.} {\it Can the result be derived from the method developed in {\rm \cite{MS}}?}

\medskip

A comment is in order. If we treat the Gaussian case
($e^{-(ax^{2}+bxy+cy^{2})}$) then the integral is reduced to
\begin{gather}
\label{eq:gaussian-formula}
\int_{{\mathbb R}}\frac{1}{ax^{2}+bx+c}dx=\frac{2\pi}{\sqrt{-D}}
\end{gather}
if $a>0$ and $D=b^{2}-4ac<0$. Noting
\begin{gather*}
  \pi=\frac{\sqrt{\pi}\sqrt{\pi}}{1}=
  \frac{\Gamma(\frac{1}{2})\Gamma(\frac{1}{2})}{\Gamma(1)}=
  B\left(\frac{1}{2},\frac{1}{2}\right)
\end{gather*}
(\ref{eq:gaussian-formula}) should be read as
\begin{gather*}
  \int_{{\mathbb R}}\frac{1}{ax^{2}+bx+c}dx=
  \frac{2B\big(\frac{1}{2},\frac{1}{2}\big)}{\sqrt{-D}}.
\end{gather*}

\section{Proof of the formula}\label{section3}

The proof is delicate.
In order to prevent possible  misunderstanding we present a detailed proof
in this section.

\begin{proof}[Proof of (I)] We prove (\ref{eq:formula-I}) in case of $D<0$.

First we consider the special case where $a=0$ in the cubic equation
$ax^{3}+bx^{2}+cx+d$. Namely, we calculate the integral
\begin{gather*}
  \int_{{\mathbb R}} \frac{1}{\sqrt[3]{(bx^{2}+cx+d)^{2}}}dx.
\end{gather*}
Noting $-D=b^{2}(4bd-c^{2})>0$ we obtain
\begin{gather}
\int_{{\mathbb R}} \frac{1}{\sqrt[3]{(bx^{2}+cx+d)^{2}}}dx
 =
\int_{{\mathbb R}} \frac{1}{\sqrt[3]{b^{2}}\sqrt[3]{(x^{2}+\frac{c}{b}x+\frac{d}{b})^{2}}}dx\nonumber\\
\qquad{}
=
\frac{1}{\sqrt[3]{b^{2}}}
\int_{{\mathbb R}}
\frac{1}{\sqrt[3]{\left((x+\frac{c}{2b})^{2}+\frac{d}{b}-\frac{c^{2}}{4b^{2}}\right)^{2}}}dx
=
\frac{1}{\sqrt[3]{b^{2}}}
\int_{{\mathbb R}}
\frac{1}{\sqrt[3]{\left(x^{2}+\frac{4bd-c^{2}}{4b^{2}}\right)^{2}}}dx
\notag \\
\qquad
\overset{T^{2}\equiv \frac{4bd-c^{2}}{4b^{2}}>0}{=}
\frac{1}{\sqrt[3]{b^{2}}}
\int_{{\mathbb R}} \frac{1}{\sqrt[3]{(x^{2}+T^{2})^{2}}}dx
%\ \Longleftarrow \ T^{2}\equiv \frac{4bd-c^{2}}{4b^{2}}>0
\overset{x=Ty}{=}
\frac{1}{\sqrt[3]{b^{2}}}
\int_{{\mathbb R}} \frac{T}{\sqrt[3]{T^{4}}\sqrt[3]{(y^{2}+1)^{2}}}dy
%\ \Longleftarrow \ x=Ty
\notag \\
\qquad{}
=
\frac{2}{\sqrt[3]{b^{2}T}}
\int_{0}^{\infty} \frac{1}{\sqrt[3]{(y^{2}+1)^{2}}}dy
\overset{y=\sqrt{x}}{=}
\frac{2}{\sqrt[3]{b^{2}T}}
\int_{0}^{\infty}\frac{1}{\sqrt[3]{(x+1)^{2}}}\frac{dx}{2\sqrt{x}}
%\ \Longleftarrow \ y=\sqrt{x}
\notag \\
\qquad{}
=
\frac{1}{\sqrt[3]{b^{2}T}}
\int_{0}^{\infty}\frac{x^{-\frac{1}{2}}}{(x+1)^{\frac{2}{3}}}dx
=
\frac{B\big(\frac{1}{2},\frac{1}{6}\big)}{\sqrt[3]{b^{2}T}}
=
\frac{B\big(\frac{1}{2},\frac{1}{6}\big)}{\sqrt[6]{b^{4}T^{2}}}
=
\frac{\sqrt[3]{2}B\big(\frac{1}{2},\frac{1}{6}\big)}{\sqrt[6]{b^{2}(4bd-c^{2})}}
=
\frac{\sqrt[3]{2}B\big(\frac{1}{2},\frac{1}{6}\big)}{\sqrt[6]{-D}}.\!\!\!\!\label{eq:f-1}
\end{gather}

Now we consider the general case of  $a\neq0$. From the condition $D<0$ there is (only) one real root
of  the cubic equation $ax^{3}+bx^{2}+cx+d=0$. Let us denote it by
$\alpha$.  From the equation
\begin{gather*}
  ax^{3}+bx^{2}+cx+d=(x-\alpha)(ax^{2}+kx+l),\qquad
  a\alpha^{3}+b\alpha^{2}+c\alpha+d=0
\end{gather*}
we have easily
\begin{gather}
  \label{eq:f-2}
  b=k-a\alpha,\qquad c=l-k\alpha,\qquad d=-l\alpha.
\end{gather}

First we assume  $\alpha=0$. In this case $d=0$ and
\begin{gather*}
  ax^{3}+bx^{2}+cx+d=x\big(ax^{2}+bx+c\big).
\end{gather*}
Then
\begin{gather*}
%\label{eq:f-3}
\int_{{\mathbb R}} \frac{1}{\sqrt[3]{x^{2}(ax^{2}+bx+c)^{2}}}dx
 =
\int_{0}^{\infty} \frac{1}{\sqrt[3]{x^{2}(ax^{2}+bx+c)^{2}}}dx
+
\int_{-\infty}^{0}\frac{1}{\sqrt[3]{x^{2}(ax^{2}+bx+c)^{2}}}dx
\notag \\
\qquad{} \overset{x=\frac{1}{y}}{=}
\int_{\infty}^{0} \frac{1}{\sqrt[3]{\frac{1}{y^{2}}\left(
\frac{a}{y^{2}}+\frac{b}{y}+c\right)^{2}}}
\left(-\frac{dy}{y^{2}}\right)
+
\int_{0}^{-\infty} \frac{1}{\sqrt[3]{\frac{1}{y^{2}}\left(
\frac{a}{y^{2}}+\frac{b}{y}+c\right)^{2}}}
\left(-\frac{dy}{y^{2}}\right)
%\ \Longleftarrow \ x=\frac{1}{y}
\notag \\
\qquad{} =
\int_{0}^{\infty} \frac{1}{\sqrt[3]{(cy^{2}+by+a)^{2}}}dy
+
\int_{-\infty}^{0} \frac{1}{\sqrt[3]{(cy^{2}+by+a)^{2}}}dy
\notag \\
\qquad{} =
\int_{{\mathbb R}}\frac{1}{\sqrt[3]{(cy^{2}+by+a)^{2}}}dy
\ \overset{\eqref{eq:f-1} \ (b\rightarrow c;\ c\rightarrow b;\ d\rightarrow a)}{=} \
\frac{\sqrt[3]{2}B\big(\frac{1}{2},\frac{1}{6}\big)}{\sqrt[6]{c^{2}(4ac-b^{2})}}
%\ \Longleftarrow \ (\ref{eq:f-1}) \ (b\rightarrow c;\ c\rightarrow b;\ d\rightarrow a)
 =
\frac{\sqrt[3]{2}B\big(\frac{1}{2},\frac{1}{6}\big)}{\sqrt[6]{-D}}.
\end{gather*}

Next, let us calculate the case $\alpha\ne 0$:
\begin{gather}
\int_{{\mathbb R}} \frac{1}{\sqrt[3]{(x-\alpha)^{2}(ax^{2}+kx+l)^{2}}}dx
\overset{x=y+\alpha}{=}
\int_{{\mathbb R}} \frac{1}{\sqrt[3]{y^{2}\{a(y+\alpha)^{2}+k(y+\alpha)+l\}^{2}}}dy
%\ \Longleftarrow \ x=y+\alpha
\notag \\
\qquad{} =
\int_{{\mathbb R}} \frac{1}{\sqrt[3]{y^{2}
\{ay^{2}+(2a\alpha+k)y+(a\alpha^{2}+k\alpha+l)\}^{2}}}dy
\notag \\
\qquad{} \overset{\eqref{eq:f-2}}{=}
\frac{\sqrt[3]{2}B\big(\frac{1}{2},\frac{1}{6}\big)}
{\sqrt[6]{(a\alpha^{2}+k\alpha+l)^{2}
\{4a(a\alpha^{2}+k\alpha+l)-(2a\alpha+k)^{2}\}}}
%\ \Longleftarrow \ (\ref{eq:f-2})
\notag \\
\qquad{} =
\frac{\sqrt[3]{2}B\big(\frac{1}{2},\frac{1}{6}\big)}
{\sqrt[6]{(a\alpha^{2}+k\alpha+l)^{2}(4al-k^{2})}}.
\label{eq:f-4}
\end{gather}

\noindent
{\bfseries Key Lemma.} {\it From \eqref{eq:f-2}
the following equation holds}
\begin{gather}
\label{eq:f-5}
\big(a\alpha^{2}+k\alpha+l\big)^{2}\big(4al-k^{2}\big)
 =
27a^{2}d^{2}+4ac^{3}-18abcd-b^{2}c^{2}+4b^{3}d
=-D.
\end{gather}

The proof is straightforward but tedious.

Therefore, from both (\ref{eq:f-4}) and (\ref{eq:f-5}) we obtain
the formula
\begin{gather*}
%\label{eq:f-6}
\int_{{\mathbb R}} \frac{1}{\sqrt[3]{(x-\alpha)^{2}(ax^{2}+kx+l)^{2}}}dx
=
\frac{\sqrt[3]{2}B\big(\frac{1}{2},\frac{1}{6}\big)}
{\sqrt[6]{-D}}.\tag*{\qed}
\end{gather*}
\renewcommand{\qed}{}
\end{proof}

\begin{proof}[Proof of (II)] We prove (\ref{eq:formula-II}) in case of $D>0$.
Let us start with the evaluation of the following  integral
\begin{gather*}
  \int_{{\mathbb R}} \frac{1}{\sqrt[3]{x^{2}(x-\alpha)^{2}}}dx
\end{gather*}
for $\alpha > 0$. Then
\begin{gather}
\int_{{\mathbb R}} \frac{1}{\sqrt[3]{x^{2}(x-\alpha)^{2}}}dx
=
\int_{-\infty}^{0}\frac{1}{\sqrt[3]{x^{2}(x-\alpha)^{2}}}dx
+
\int_{0}^{\infty}\frac{1}{\sqrt[3]{x^{2}(x-\alpha)^{2}}}dx
\notag \\
\hphantom{\int_{{\mathbb R}} \frac{1}{\sqrt[3]{x^{2}(x-\alpha)^{2}}}dx}{}
=
\int_{0}^{\infty}\frac{1}{\sqrt[3]{x^{2}(x+\alpha)^{2}}}dx
+
\int_{0}^{\infty}\frac{1}{\sqrt[3]{x^{2}(x-\alpha)^{2}}}dx,\label{eq:f-7}
\end{gather}
where the change of variable $x\rightarrow -x$ for the f\/irst term
of the right hand side was made.

Each term can be evaluated elementarily:
\begin{gather*}
\int_{0}^{\infty}\!\!\frac{1}{\sqrt[3]{x^{2}(x+\alpha)^{2}}}dx
 \overset{x=\alpha t}{=}
\frac{1}{\sqrt[3]{\alpha}}
\int_{0}^{\infty}\!\!\frac{1}{\sqrt[3]{t^{2}(t+1)^{2}}}dt
%\ \Longleftarrow x=\alpha t
 =
\alpha^{-\frac{1}{3}}
\int_{0}^{\infty}\!\! \frac{t^{-\frac{2}{3}}}{(t+1)^{\frac{2}{3}}}dt
=
\alpha^{-\frac{1}{3}}B\left(\frac{1}{3},\frac{1}{3}\right),
\end{gather*}
while
\begin{gather*}
\int_{0}^{\infty}\frac{1}{\sqrt[3]{x^{2}(x-\alpha)^{2}}}dx
\overset{x=\alpha t}{=}
\alpha^{-\frac{1}{3}}
\int_{0}^{\infty}\frac{1}{\sqrt[3]{t^{2}(t-1)^{2}}}dt
% \Longleftarrow x=\alpha t
\\
\hphantom{\int_{0}^{\infty}\frac{1}{\sqrt[3]{x^{2}(x-\alpha)^{2}}}dx}{} =
\alpha^{-\frac{1}{3}}
\left\{
\int_{0}^{1}\frac{1}{\sqrt[3]{t^{2}(t-1)^{2}}}dt
+
\int_{1}^{\infty}\frac{1}{\sqrt[3]{t^{2}(t-1)^{2}}}dt
\right\}  \\
\hphantom{\int_{0}^{\infty}\frac{1}{\sqrt[3]{x^{2}(x-\alpha)^{2}}}dx}{}
=
\alpha^{-\frac{1}{3}}
\left\{
\int_{0}^{1}\frac{1}{\sqrt[3]{t^{2}(1-t)^{2}}}dt
+
\int_{1}^{\infty}\frac{1}{\sqrt[3]{t^{2}(t-1)^{2}}}dt
\right\}  \\
\hphantom{\int_{0}^{\infty}\frac{1}{\sqrt[3]{x^{2}(x-\alpha)^{2}}}dx}{}
=
2\alpha^{-\frac{1}{3}}
\int_{0}^{1}\frac{1}{\sqrt[3]{t^{2}(1-t)^{2}}}dt
=
2\alpha^{-\frac{1}{3}}
\int_{0}^{1}t^{-\frac{2}{3}}(1-t)^{-\frac{2}{3}}dt  \\
\hphantom{\int_{0}^{\infty}\frac{1}{\sqrt[3]{x^{2}(x-\alpha)^{2}}}dx}{}
=
2\alpha^{-\frac{1}{3}}B\left(\frac{1}{3},\frac{1}{3}\right),
\end{gather*}
where we have used
\begin{gather*}
\int_{1}^{\infty}\!\frac{1}{\sqrt[3]{t^{2}(t-1)^{2}}}dt
\overset{t=\frac{1}{s}}{=}
\int_{1}^{0}\! \frac{1}{\sqrt[3]{\frac{1}{s^{2}}\frac{(1-s)^{2}}{s^{2}}}}\left(-\frac{ds}{s^{2}}\right)
%\Longleftarrow t=\frac{1}{s}
 =
\int_{0}^{1}\!\frac{1}{\sqrt[3]{s^{2}(1-s)^{2}}}ds
=
\int_{0}^{1}\!\frac{1}{\sqrt[3]{t^{2}(1-t)^{2}}}dt.
\end{gather*}
From (\ref{eq:f-7}) we have
\begin{gather*}
%\label{eq:f-7fin}% relabeled
\int_{{\mathbb R}} \frac{1}{\sqrt[3]{x^{2}(x-\alpha)^{2}}}dx
=
3\alpha^{-\frac{1}{3}}B\left(\frac{1}{3},\frac{1}{3}\right)
=
\frac{3B\big(\frac{1}{3},\frac{1}{3}\big)}{\sqrt[3]{\alpha}}.
\end{gather*}

Now we consider the special  case $a=0$ in the cubic equation
$ax^{3}+bx^{2}+cx+d$. Then by $D=b^{2}(c^{2}-4bd)>0$
we obtain
\begin{gather}
\int_{{\mathbb R}} \frac{1}{\sqrt[3]{(bx^{2}+cx+d)^{2}}}dx
=
\int_{-\infty}^{\infty}\frac{1}{\sqrt[3]{
\left\{b(x+\frac{c}{2b})^{2}-\frac{c^{2}-4bd}{4b}\right\}^{2}
}}dx
=
\int_{-\infty}^{\infty}\frac{1}{\sqrt[3]{
(bx^{2}-\frac{c^{2}-4bd}{4b})^{2}
}}dx
\notag \\
\qquad{} \overset{\alpha^{2}=\frac{c^{2}-4bd}{4b^{2}}>0}{=}
\frac{1}{\sqrt[3]{b^{2}}}
\int_{-\infty}^{\infty}\frac{1}{\sqrt[3]{
(x^{2}-\alpha^{2})^{2}
}}dx
%\ \Longleftarrow \alpha^{2}=\frac{c^{2}-4bd}{4b^{2}}>0
%\notag \\
=
\frac{1}{\sqrt[3]{b^{2}}}
\int_{-\infty}^{\infty}\frac{1}{\sqrt[3]{
(x-\alpha)^{2}(x+\alpha)^{2}}}dx
\notag \\
\qquad{} \overset{y=x+\alpha}{=}
\frac{1}{\sqrt[3]{b^{2}}}
\int_{-\infty}^{\infty}\frac{1}{\sqrt[3]{
y^{2}(y-2\alpha)^{2}}}dy
%\ \Longleftarrow y=x+\alpha
%\notag \\
\overset{\eqref{eq:f-7}}{=}
\frac{1}{\sqrt[3]{b^{2}}}\frac{3B\big(\frac{1}{3},\frac{1}{3}\big)}{\sqrt[3]{2\alpha}}
%\ \Longleftarrow (\ref{eq:f-7})
\notag \\
\qquad{} =
\frac{3B\big(\frac{1}{3},\frac{1}{3}\big)}{\sqrt[3]{2\alpha b^{2}}}
=
\frac{3B\big(\frac{1}{3},\frac{1}{3}\big)}{\sqrt[6]{4\alpha^{2}b^{4}}}
=
\frac{3B\big(\frac{1}{3},\frac{1}{3}\big)}{\sqrt[6]{b^{2}(c^{2}-4bd)}}
=
\frac{3B\big(\frac{1}{3},\frac{1}{3}\big)}{\sqrt[6]{D}}.\label{eq:f-8}
\end{gather}

Next we consider the remaining general case of $a\neq0$. From the condition $D>0$ there are three real solutions
in the equation $ax^{3}+bx^{2}+cx+d=0$. We denote one of them by
$\alpha$.  Remember the relations
$b=k-a\alpha$, $c=l-k\alpha$, $d=-l\alpha$
from the equation
\begin{gather*}
  ax^{3}+bx^{2}+cx+d=(x-\alpha)\big(ax^{2}+kx+l\big),\qquad
  a\alpha^{3}+b\alpha^{2}+c\alpha+d=0.
\end{gather*}

First we assume $\alpha=0$. Then
\begin{gather*}
  ax^{3}+bx^{2}+cx+d=x(ax^{2}+bx+c)
\end{gather*}
and from $D=c^{2}(b^{2}-4ac)$ we have
\begin{gather}
\int_{{\mathbb R}} \frac{1}{\sqrt[3]{x^{2}(ax^{2}+bx+c)^{2}}}dx
=
\int_{0}^{\infty} \frac{1}{\sqrt[3]{x^{2}(ax^{2}+bx+c)^{2}}}dx
+
\int_{-\infty}^{0}\frac{1}{\sqrt[3]{x^{2}(ax^{2}+bx+c)^{2}}}dx
\notag \\
\qquad{} \overset{x=\frac{1}{y}}{=}
\int_{\infty}^{0} \frac{1}{\sqrt[3]{\frac{1}{y^{2}}\left(
\frac{a}{y^{2}}+\frac{b}{y}+c\right)^{2}}}
\left(-\frac{dy}{y^{2}}\right)
+
\int_{0}^{-\infty} \frac{1}{\sqrt[3]{\frac{1}{y^{2}}\left(
\frac{a}{y^{2}}+\frac{b}{y}+c\right)^{2}}}
\left(-\frac{dy}{y^{2}}\right)
%\ \Longleftarrow \ x=\frac{1}{y}
\notag \\
\qquad{} =
\int_{0}^{\infty} \frac{1}{\sqrt[3]{(cy^{2}+by+a)^{2}}}dy
+
\int_{-\infty}^{0} \frac{1}{\sqrt[3]{(cy^{2}+by+a)^{2}}}dy
=
\int_{{\mathbb R}} \frac{1}{\sqrt[3]{(cy^{2}+by+a)^{2}}}dy
\notag \\
\qquad{}
\overset{\eqref{eq:f-8} \ (c\rightarrow b;\ b\rightarrow c;\ a\rightarrow d)}{=}
\frac{3B\big(\frac{1}{3},\frac{1}{3}\big)}{\sqrt[6]{c^{2}(b^{2}-4ac)}}
%\ \Longleftarrow \ (\ref{eq:f-8}) \ (c\rightarrow b;\ b\rightarrow c;\ a\rightarrow d)
 =
\frac{3B\big(\frac{1}{3},\frac{1}{3}\big)}{\sqrt[6]{D}}.\label{eq:f-9}
\end{gather}

For the case $\alpha\ne 0$ we obtain the formula
\begin{gather*}
%\label{eq:f-10}
\int_{{\mathbb R}} \frac{1}{\sqrt[3]{(x-\alpha)^{2}(ax^{2}+kx+l)^{2}}}dx
\overset{x=y+\alpha}{=}
\int_{{\mathbb R}} \frac{1}{\sqrt[3]{y^{2}\{a(y+\alpha)^{2}+k(y+\alpha)+l\}^{2}}}dy
%\ \Longleftarrow \ x=y+\alpha
\\
\qquad {} =
\int_{{\mathbb R}} \frac{1}{\sqrt[3]{y^{2}
\{ay^{2}+(2a\alpha+k)y+(a\alpha^{2}+k\alpha+l)\}^{2}}}dy
\notag \\
\qquad{} \overset{\eqref{eq:f-9}}{=}
\frac{3B\big(\frac{1}{3},\frac{1}{3}\big)}
{\sqrt[6]{(a\alpha^{2}+k\alpha+l)^{2}
\{(2a\alpha+k)^{2}-4a(a\alpha^{2}+k\alpha+l)\}}}
%\ \Longleftarrow \ (\ref{eq:f-9})
\notag \\
\qquad{} =
\frac{3B\big(\frac{1}{3},\frac{1}{3}\big)}
{\sqrt[6]{(a\alpha^{2}+k\alpha+l)^{2}(k^{2}-4al)}}
\overset{\eqref{eq:f-5}}{=}
\frac{3B\big(\frac{1}{3},\frac{1}{3}\big)}{\sqrt[6]{D}}.\tag*{\qed}
%\ \Longleftarrow \  (\ref{eq:f-5})
\end{gather*}
\renewcommand{\qed}{}
\end{proof}

\begin{proof}[Proof of (III)] We prove the relation (\ref{eq:formula-III}).
Let us make some preparations.  For the Gamma function
(\ref{eq:Gamma-function}) there are well-known formulas
(see for example \cite{WW})
\begin{gather}
\label{eq:f-11}
B(x,y)=\frac{\Gamma(x)\Gamma(y)}{\Gamma(x+y)}\qquad (x,y>0), \\
\label{eq:f-12}
\Gamma(x)\Gamma(1-x)=\frac{\pi}{\sin(\pi x)}\qquad (0<x<1), \\
\label{eq:f-13}
\Gamma\left(\frac{x}{2}\right)\Gamma\left(\frac{x+1}{2}\right)
=
\frac{\sqrt{\pi}}{2^{x-1}}\Gamma(x)
=
2^{1-x}\Gamma\left(\frac{1}{2}\right)\Gamma(x).
\end{gather}

(\ref{eq:f-13}) is called the Legendre's relation. In the formula we set
$x=2/3$, then
\begin{gather*}
  \Gamma\left(\frac{1}{3}\right)\Gamma\left(\frac{5}{6}\right)
  =
  \sqrt[3]{2}\Gamma\left(\frac{1}{2}\right)\Gamma\left(\frac{2}{3}\right).
\end{gather*}
Multiplying both sides by $\Gamma(1/6)$ gives
\begin{gather*}
 \Gamma\left(\frac{1}{3}\right)\Gamma\left(\frac{5}{6}\right)\Gamma\left(\frac{1}{6}\right)
=
\sqrt[3]{2}\Gamma\left(\frac{1}{2}\right)\Gamma\left(\frac{2}{3}\right)
\Gamma\left(\frac{1}{6}\right)  \\
\qquad{} \Longleftrightarrow
\quad\Gamma\left(\frac{1}{3}\right)\frac{\pi}{\sin(\frac{\pi}{6})}
=
\sqrt[3]{2}\Gamma\left(\frac{1}{2}\right)\Gamma\left(\frac{1}{6}\right)
\Gamma\left(\frac{2}{3}\right)
\\
\qquad{} \Longleftrightarrow
\quad 2\pi\Gamma\left(\frac{1}{3}\right)
=
\sqrt[3]{2}\Gamma\left(\frac{1}{2}\right)\Gamma\left(\frac{1}{6}\right)
\Gamma\left(\frac{2}{3}\right) \\
\qquad{} \Longleftrightarrow
\quad2\pi
\frac{\Gamma\left(\frac{1}{3}\right)}{\Gamma\left(\frac{2}{3}\right)^{2}}
=
\sqrt[3]{2}
\frac{\Gamma\left(\frac{1}{2}\right)\Gamma\left(\frac{1}{6}\right)}
{\Gamma\left(\frac{2}{3}\right)} \\
\qquad{} \Longleftrightarrow
\quad2\pi
\frac{\Gamma\left(\frac{1}{3}\right)\Gamma\left(\frac{1}{3}\right)}
{\Gamma\left(\frac{2}{3}\right)\Gamma\left(\frac{1}{3}\right)
\Gamma\left(\frac{2}{3}\right)}
=
\sqrt[3]{2}B\left(\frac{1}{2},\frac{1}{6}\right)
\\
\qquad{} \Longleftrightarrow
\quad2\pi
\frac{\Gamma\left(\frac{1}{3}\right)^{2}}
{\frac{\pi}{\sin (\frac{\pi}{3})}\Gamma\left(\frac{2}{3}\right)}
=
\sqrt[3]{2}B\left(\frac{1}{2},\frac{1}{6}\right)
\\
\qquad{} \Longleftrightarrow
\quad\sqrt{3}B\left(\frac{1}{3},\frac{1}{3}\right)
=
\sqrt[3]{2}B\left(\frac{1}{2},\frac{1}{6}\right),
\end{gather*}
where we have used formulas (\ref{eq:f-11}) and
 (\ref{eq:f-12})  several times.

The proof of (\ref{eq:formula-III}) is now complete.
\end{proof}

\section{Renormalized integral revisited}\label{section4}

In this section let us check whether the renormalized integral
(\ref{eq:renorm-definition}) is reasonable or not
by making use of the results in the Section~\ref{section3}.

In the introduction we introduced the following integral def\/ined on
$D_{R}=[-R,R]\times [-R,R]$
\begin{gather*}
  F(a,b,c,d)=\iint_{D_{R}}e^{-\left(ax^{3}+bx^{2}y+cxy^{2}+dy^{3}\right)}dxdy.
\end{gather*}
For this it is easy to see
\begin{gather}
\left(
\frac{\partial}{\partial a}\frac{\partial}{\partial d}-
\frac{\partial}{\partial b}\frac{\partial}{\partial c}
\right)F(a,b,c,d)\notag\\
\qquad{} =
\iint_{D_{R}}\big(x^{3}\cdot y^{3}-x^{2}y\cdot xy^{2}\big)
e^{-\left(ax^{3}+bx^{2}y+cxy^{2}+dy^{3}\right)}dxdy
=0, \notag \\
\left(
\frac{\partial}{\partial b}\frac{\partial}{\partial b}-
\frac{\partial}{\partial a}\frac{\partial}{\partial c}
\right)F(a,b,c,d)\notag\\
\qquad{}
=
\iint_{D_{R}}\big(x^{2}y\cdot x^{2}y-x^{3}\cdot xy^{2}\big)
e^{-\left(ax^{3}+bx^{2}y+cxy^{2}+dy^{3}\right)}dxdy
=0,  \notag \\
\left(
\frac{\partial}{\partial c}\frac{\partial}{\partial c}-
\frac{\partial}{\partial b}\frac{\partial}{\partial d}
\right)F(a,b,c,d)\notag\\
\qquad{}
=
\iint_{D_{R}}\big(xy^{2}\cdot xy^{2}-x^{2}y\cdot y^{3}\big)
e^{-\left(ax^{3}+bx^{2}y+cxy^{2}+dy^{3}\right)}dxdy
=0.  \label{eq:identity-1}
\end{gather}

On the other hand, if we set
\begin{gather*}
  {\cal F}(a,b,c,d)
  =\int_{{\mathbb R}} \frac{1}{\sqrt[3]{(ax^{3}+bx^{2}+cx+d)^{2}}}dx
  =\frac{C_{\pm}}{\sqrt[6]{\pm D}},
\end{gather*}
then we can also verify the same relations:
\begin{gather}
\left(
\frac{\partial}{\partial a}\frac{\partial}{\partial d}-
\frac{\partial}{\partial b}\frac{\partial}{\partial c}
\right){\cal F}(a,b,c,d)=0, \qquad
\left(
\frac{\partial}{\partial b}\frac{\partial}{\partial b}-
\frac{\partial}{\partial a}\frac{\partial}{\partial c}
\right){\cal F}(a,b,c,d)=0, \notag\\
\left(
\frac{\partial}{\partial c}\frac{\partial}{\partial c}-
\frac{\partial}{\partial b}\frac{\partial}{\partial d}
\right){\cal F}(a,b,c,d)=0. \label{eq:identity-2}
\end{gather}
Verif\/ication by hand is rather tough, but it can be done easily
by use of MATHEMATICA\footnote{The author owes
the calculation to Hiroshi Oike.}.

From (\ref{eq:identity-1}) and (\ref{eq:identity-2}) we can conclude
that our renormalized integral  (\ref {eq:renorm-definition})
is reasonable enough.

\section{Discriminant}\label{section5}
In this section we make some comments on the discriminant
(\ref{eq:cubic-discriminant}). See \cite{Sa} for  more details
(\cite{Sa} is strongly recommended).

For the equations
\begin{gather}
\label{eq:cubic-function}
f(x)=ax^{3}+bx^{2}+cx+d,\qquad
f^{\prime}(x)=3ax^{2}+2bx+c
\end{gather}
the resultant $R(f,f^{\prime})$ of $f$ and $f^{\prime}$ is given by
\begin{gather}
\label{eq:resultant}
R(f,f^{\prime})=
\left|
  \begin{array}{ccccc}
   a  & b  & c  & d  & 0  \\
   0  & a  & b  & c  & d  \\
   3a & 2b & c  & 0  & 0  \\
   0  & 3a & 2b & c  & 0  \\
   0  & 0  & 3a & 2b & c
  \end{array}
\right|.
\end{gather}
It is easy to calculate (\ref{eq:resultant}) and the result becomes
\begin{gather*}
%\label{eq:}
\frac{1}{a}R(f,f^{\prime})
=27a^{2}d^{2}+4ac^{3}-18abcd-b^{2}c^{2}+4b^{3}d
=-D.
\end{gather*}

On the other hand, if $\alpha$, $\beta$, $\gamma$ are three solutions
of $f(x)=0$ in (\ref{eq:cubic-function}), then the following relations
are well--known
\begin{gather*}
\alpha +\beta +\gamma =-\frac{b}{a},  \qquad
\alpha\beta +\alpha\gamma +\beta\gamma =\frac{c}{a}, \qquad
\alpha\beta\gamma =-\frac{d}{a}.
\end{gather*}
From these  it is easy to see
\begin{gather*}
\alpha +\beta +\gamma =-\frac{b}{a},  \qquad
\alpha^{2} +\beta^{2} +\gamma^{2} =\frac{b^{2}-2ac}{a^{2}},  \qquad
\alpha^{3} +\beta^{3} +\gamma^{3} =-\frac{b^{3}+3a^{2}d-3abc}{a^{3}},  \\
\alpha^{4} +\beta^{4} +\gamma^{4} =\frac{b^{4}+4a^{2}bd+2a^{2}c^{2}-4ab^{2}c}{a^{4}}.
\end{gather*}

If we set
\begin{gather*}
%\label{eq:Delta}
\Delta =(\alpha-\beta)(\alpha-\gamma)(\beta-\gamma)
\end{gather*}
the discriminant $D$ is given by
\begin{gather*}
%\label{eq:definition}
D=a^{4}\Delta^{2}.
\end{gather*}
Let us calculate $\Delta^{2}$ directly. For the Vandermonde matrix
\begin{gather*}
V=
\left(
  \begin{array}{ccc}
   1          & 1         & 1           \\
   \alpha     & \beta     & \gamma      \\
   \alpha^{2} & \beta^{2} & \gamma^{2}
  \end{array}
\right)\quad
\Longrightarrow\quad
|V|=-\Delta
\end{gather*}
we obtain by some manipulations of determinant
\begin{gather*}
\Delta^{2}
=(-|V|)^{2}=|V||V^{T}|=|VV^{T}|
=
\left|
  \begin{array}{ccc}
   3 & \alpha +\beta +\gamma & \alpha^{2}+\beta^{2}+\gamma^{2} \\
   \alpha +\beta +\gamma & \alpha^{2}+\beta^{2}+\gamma^{2} &
   \alpha^{3}+\beta^{3}+\gamma^{3}  \\
   \alpha^{2}+\beta^{2}+\gamma^{2} & \alpha^{3}+\beta^{3}+\gamma^{3} &
   \alpha^{4}+\beta^{4}+\gamma^{4}
  \end{array}
\right|     \\
\hphantom{\Delta^{2}}{} =
\left|
  \begin{array}{ccc}
   3 & -\dfrac{b}{a} & \dfrac{b^{2}-2ac}{a^{2}}  \vspace{2mm}\\
  -\dfrac{b}{a} & \dfrac{b^{2}-2ac}{a^{2}} & -\dfrac{b^{3}+3a^{2}d-3abc}{a^{3}}  \vspace{2mm}\\
  \dfrac{b^{2}-2ac}{a^{2}} & -\dfrac{b^{3}+3a^{2}d-3abc}{a^{3}} &
  \dfrac{b^{4}+4a^{2}bd+2a^{2}c^{2}-4ab^{2}c}{a^{4}}
  \end{array}
\right|   \\
\hphantom{\Delta^{2}}{}
=
\left|
  \begin{array}{ccc}
   3 & -\dfrac{b}{a} & \dfrac{b^{2}-2ac}{a^{2}}  \vspace{2mm}\\
   \dfrac{2b}{a} & -\dfrac{2c}{a} & -\dfrac{3ad-bc}{a^{2}}  \vspace{2mm}\\
  -\dfrac{2b^{2}+2ac}{a^{2}} & -\dfrac{3ad-3bc}{a^{2}} & \dfrac{4abd+2ac^{2}-2b^{2}c}{a^{3}}
  \end{array}
\right|  \\
\hphantom{\Delta^{2}}{}
=
\left|
  \begin{array}{ccc}
   3 & -\dfrac{b}{a} & \dfrac{b^{2}-2ac}{a^{2}}  \vspace{2mm}\\
   \dfrac{2b}{a} & -\dfrac{2c}{a} & -\dfrac{3ad-bc}{a^{2}}  \vspace{2mm}\\
  -\dfrac{2c}{a} & -\dfrac{3ad-bc}{a^{2}} & \dfrac{abd+2ac^{2}-b^{2}c}{a^{3}}
  \end{array}
\right|
=
\left|
  \begin{array}{ccc}
   3 & -\dfrac{b}{a} & -\dfrac{2c}{a}  \vspace{2mm}\\
   \dfrac{2b}{a} & -\dfrac{2c}{a} & -\dfrac{3ad+bc}{a^{2}} \vspace{2mm} \\
  -\dfrac{2c}{a} & -\dfrac{3ad-bc}{a^{2}} & \dfrac{-2bd+2c^{2}}{a^{2}}
  \end{array}
\right|   \\
\hphantom{\Delta^{2}}{}
=
\left|
  \begin{array}{ccc}
   3 & -\dfrac{b}{a} & -\dfrac{2c}{a}  \vspace{2mm}\\
   0 & 2\dfrac{b^{2}-3ac}{3a^{2}} & \dfrac{bc-9ad}{3a^{2}} \vspace{2mm}\\
   0 & \dfrac{bc-9ad}{3a^{2}} & 2\dfrac{c^{2}-3bd}{3a^{2}}
  \end{array}
\right|
=
3
\left|
  \begin{array}{cc}
   2\dfrac{b^{2}-3ac}{3a^{2}} & \dfrac{bc-9ad}{3a^{2}} \vspace{2mm}\\
   \dfrac{bc-9ad}{3a^{2}} & 2\dfrac{c^{2}-3bd}{3a^{2}}
  \end{array}
\right|   \\
\hphantom{\Delta^{2}}{}
=\frac{1}{a^{4}}
\frac{-1}{3}\left\{(bc-9ad)^{2}-4(b^{2}-3ac)(c^{2}-3bd)\right\}.
\end{gather*}

This result is very suggestive. In fact, from the cubic equation
\begin{gather*}
  ax^{3}+bx^{2}+cx+d=0
\end{gather*}
we have three data
\begin{gather*}
  A=b^{2}-3ac,\qquad B=bc-9ad,\qquad C=c^{2}-3bd,
\end{gather*}
 and so if we consider the quadratic equation
\begin{gather*}
  AX^{2}+BX+C=0
\end{gather*}
then the discriminant is just $B^{2}-4AC$. This is very interesting.

\medskip

\noindent
{\bfseries Problem.} {\it Clarify the above connection.}

\medskip

As a result we have
\begin{gather*}
D
=\frac{-1}{3}\left\{(bc-9ad)^{2}-4(b^{2}-3ac)(c^{2}-3bd)\right\}
=b^{2}c^{2}+18abcd-4ac^{3}-4b^{3}d-27a^{2}d^{2}.
\end{gather*}

\section{Some calculations}\label{section6}

In this section we calculate some quantities coming from
the integral.

The expectation value $\langle{x^{3}}\rangle$ is formally given by
\begin{gather*}
\langle{x^{3}}\rangle  =
\frac{\displaystyle \iint x^{3}e^{-\left(ax^{3}+bx^{2}y+cxy^{2}+dy^{3}\right)}dxdy}
{\displaystyle \iint e^{-\left(ax^{3}+bx^{2}y+cxy^{2}+dy^{3}\right)}dxdy}
%\\
% \phantom{\langle{x^{3}}\rangle}{}
=
-\frac{\partial}{\partial a}\log
\left\{\iint e^{-\left(ax^{3}+bx^{2}y+cxy^{2}+dy^{3}\right)}dxdy\right\},
\end{gather*}
so renormalized expectation values
$\langle{x^{3}}\rangle_{\rm RN}$, $\langle{x^{2}y}\rangle_{\rm RN}$,
$\langle{xy^{2}}\rangle_{\rm RN}$, $\langle{y^{3}}\rangle_{\rm RN}$
are def\/ined as

\begin{definition}
\begin{gather*}
\langle{x^{3}}\rangle_{\rm RN}
=
-\frac{\partial}{\partial a}\log
\left\{
\ddagger \iint_{{\mathbb R}^{2}}e^{-(ax^{3}+bx^{2}y+cxy^{2}+dy^{3})}dxdy\ \ddagger
\right\},   \\
\langle{x^{2}y}\rangle_{\rm RN}
=
-\frac{\partial}{\partial b}\log
\left\{
\ddagger \iint_{{\mathbb R}^{2}}e^{-(ax^{3}+bx^{2}y+cxy^{2}+dy^{3})}dxdy\ \ddagger
\right\},   \\
\langle{xy^{2}}\rangle_{\rm RN}
=
-\frac{\partial}{\partial c}\log
\left\{
\ddagger \iint_{{\mathbb R}^{2}}e^{-(ax^{3}+bx^{2}y+cxy^{2}+dy^{3})}dxdy\ \ddagger
\right\},   \\
\langle{y^{3}}\rangle_{\rm RN}
=
-\frac{\partial}{\partial d}\log
\left\{
\ddagger \iint_{{\mathbb R}^{2}}e^{-(ax^{3}+bx^{2}y+cxy^{2}+dy^{3})}dxdy\ \ddagger
\right\}.
\end{gather*}
\end{definition}

From the integral forms (\ref{eq:formula-I}) and (\ref{eq:formula-II})
it is easy to calculate the above. Namely, we have
\begin{alignat*}{3}
& \langle{x^{3}}\rangle_{\rm RN} =
\frac{18bcd-4c^{3}-54ad^{2}}{6D},   \qquad && \langle{x^{2}y}\rangle_{\rm RN} =
\frac{2bc^{2}+18acd-12b^{2}d}{6D},  &  \\
& \langle{xy^{2}}\rangle_{\rm RN} = \frac{2b^{2}c+18abd-12ac^{2}}{6D},  \qquad &&
\langle{y^{3}}\rangle_{\rm RN} = \frac{18abc-4b^{3}-54a^{2}d}{6D}, &
\end{alignat*}
where $D=b^{2}c^{2}+18abcd-4ac^{3}-4b^{3}d-27a^{2}d^{2}$.

We can calculate other quantities like $\langle{x^{5}y}\rangle_{\rm RN}$
or $\langle{x^{4}y^{2}}\rangle_{\rm RN}$ by use of these ones,
which will be left to readers.

\section{Concluding remarks}\label{section7}

In this paper we calculated the non-Gaussian integral
(\ref{eq:cubic-integral}) in a direct manner and, moreover, calculated
some renormalized expectation values.  It is not clear at the present
time whether these results are useful enough or not.
It would be desirable to accumulate many supporting evidences.
Some application(s) will be reported elsewhere \cite{Fu2}.

At this stage we can consider a further generalization. Namely, for the general
degree $n$ polynomial
\begin{gather*}
%\label{eq:general-equation}
f(x)=a_{0}x^{n}+a_{1}x^{n-1}+\cdots +a_{n-1}x+a_{n}
\end{gather*}
the (non-Gaussian) integral becomes
\begin{gather}
\label{eq:general-integral}
\int_{{\mathbb R}}\frac{1}{\sqrt[n]{f(x)^{2}}}dx.
\end{gather}

The discriminant $D$ of the equation $f(x)=0$ is given by
the resultant $R(f,f^{\prime})$ of $f$ and $f^{\prime}$ like
\begin{gather*}
\frac{1}{a_{0}}R(f,f^{\prime})=(-1)^{\frac{n(n-1)}{2}}D
\ \Longleftrightarrow \
D=(-1)^{\frac{n(n-1)}{2}}R(f,f^{\prime})/a_{0}
\end{gather*}
where
\begin{gather*}
  R(f,f^{\prime})=
  \left|
    \begin{array}{cccccccc}
      a_{0} & a_{1} & \cdots & a_{n-1} & a_{n} &  &  &       \\
      & a_{0} & a_{1} & \cdots & a_{n-1} & a_{n} &  &     \\
      &  & \ddots &    & & \ddots  &   &                  \\
      &  & & a_{0} & a_{1} & \cdots & a_{n-1} & a_{n}     \\
      na_{0} & (n-1)a_{1} & \cdots & a_{n-1} &    & & &      \\
      & na_{0} & (n-1)a_{1} & \cdots & a_{n-1} & & &      \\
      &  & \ddots &   &   & \ddots   &   &                \\
      & & & & na_{0} & (n-1)a_{1} & \cdots & a_{n-1}
    \end{array}
  \right|,
\end{gather*}
see (\ref{eq:cubic-function}) and (\ref{eq:resultant}).

For example, for $n=4$
\begin{gather*}
R(f,f^{\prime})=
\left|
  \begin{array}{ccccccc}
  a_{0} & a_{1} & a_{2} & a_{3} & a_{4} & 0 & 0    \\
  0 & a_{0} & a_{1} & a_{2} & a_{3} & a_{4} & 0    \\
  0 & 0 & a_{0} & a_{1} & a_{2} & a_{3} & a_{4}    \\
  4a_{0} & 3a_{1} & 2a_{2} & a_{3} & 0 & 0 & 0   \\
  0 & 4a_{0} & 3a_{1} & 2a_{2} & a_{3} & 0 & 0   \\
  0 & 0 & 4a_{0} & 3a_{1} & 2a_{2} & a_{3} & 0   \\
  0 & 0 & 0 & 4a_{0} & 3a_{1} & 2a_{2} & a_{3}
  \end{array}
\right|
\end{gather*}
and
\begin{gather*}
D =256a_{0}^{3}a_{4}^{3}-4a_{1}^{3}a_{3}^{3}
-27a_{0}^{2}a_{3}^{4}-27a_{1}^{4}a_{4}^{2}
-128a_{0}^{2}a_{2}^{2}a_{4}^{2}
+a_{1}^{2}a_{2}^{2}a_{3}^{2}+16a_{0}a_{2}^{4}a_{4}   \\
\phantom{D=}{}
-4a_{0}a_{2}^{3}a_{3}^{2}-4a_{1}^{2}a_{2}^{3}a_{4}
+144a_{0}^{2}a_{2}a_{3}^{2}a_{4}-6a_{0}a_{1}^{2}a_{3}^{2}a_{4}
+144a_{0}a_{1}^{2}a_{2}a_{4}^{2}
-192a_{0}^{2}a_{1}a_{3}a_{4}^{2}  \\
\phantom{D=}{}
+18a_{0}a_{1}a_{2}a_{3}^{3}
+18a_{1}^{3}a_{2}a_{3}a_{4}-80a_{0}a_{1}a_{2}^{2}a_{3}a_{4},
\end{gather*}
and for $n=5$
\begin{gather*}
R(f,f^{\prime})=
\left|
  \begin{array}{ccccccccc}
  a_{0} & a_{1} & a_{2} & a_{3} & a_{4} & a_{5} & 0 & 0 & 0   \\
  0 & a_{0} & a_{1} & a_{2} & a_{3} & a_{4} & a_{5} & 0 & 0   \\
  0 & 0 & a_{0} & a_{1} & a_{2} & a_{3} & a_{4} & a_{5}  & 0  \\
  0 & 0 & 0 & a_{0} & a_{1} & a_{2} & a_{3} & a_{4} & a_{5}    \\
  5a_{0} & 4a_{1} & 3a_{2} & 2a_{3} & a_{4} & 0 & 0 & 0 & 0  \\
  0 & 5a_{0} & 4a_{1} & 3a_{2} & 2a_{3} & a_{4} & 0 & 0 & 0  \\
  0 & 0 & 5a_{0} & 4a_{1} & 3a_{2} & 2a_{3} & a_{4} & 0 & 0  \\
  0 & 0 & 0 & 5a_{0} & 4a_{1} & 3a_{2} & 2a_{3} & a_{4} & 0  \\
  0 & 0 & 0 & 0 & 5a_{0} & 4a_{1} & 3a_{2} & 2a_{3} & a_{4}
  \end{array}
\right|
\end{gather*}
and
\begin{gather*}
D =3125a_{0}^{4}a_{5}^{4}-2500a_{0}^{3}a_{1}a_{4}a_{5}^{3}
-3750a_{0}^{3}a_{2}a_{3}a_{5}^{3}+2000a_{0}^{3}a_{2}a_{4}^{2}a_{5}^{2}
+2250a_{0}^{3}a_{3}^{2}a_{4}a_{5}^{2} \\
\phantom{D=}{}
-1600a_{0}^{3}a_{3}a_{4}^{3}a_{5}+256a_{0}^{3}a_{4}^{5}
+2000a_{0}^{2}a_{1}^{2}a_{3}a_{5}^{3}-50a_{0}^{2}a_{1}^{2}a_{4}^{2}a_{5}^{2}
+2250a_{0}^{2}a_{1}a_{2}^{2}a_{5}^{3} \\
\phantom{D=}{}
-2050a_{0}^{2}a_{1}a_{2}a_{3}a_{4}a_{5}^{2}+160a_{0}^{2}a_{1}a_{2}a_{4}^{3}a_{5}
-900a_{0}^{2}a_{1}a_{3}^{3}a_{5}^{2}+1020a_{0}^{2}a_{1}a_{3}^{2}a_{4}^{2}a_{5}
 \\
\phantom{D=}{}
-192a_{0}^{2}a_{1}a_{3}a_{4}^{4}-900a_{0}^{2}a_{2}^{3}a_{4}a_{5}^{2}+825a_{0}^{2}a_{2}^{2}a_{3}^{2}a_{5}^{2}
+560a_{0}^{2}a_{2}^{2}a_{3}a_{4}^{2}a_{5}-128a_{0}^{2}a_{2}^{2}a_{4}^{4}
 \\
\phantom{D=}{}
-630a_{0}^{2}a_{2}a_{3}^{3}a_{4}a_{5}+144a_{0}^{2}a_{2}a_{3}^{2}a_{4}^{3}+108a_{0}^{2}a_{3}^{5}a_{5}
-27a_{0}^{2}a_{3}^{4}a_{4}^{2}-1600a_{0}a_{1}^{3}a_{2}a_{5}^{3}
 \\
\phantom{D=}{}
+160a_{0}a_{1}^{3}a_{3}a_{4}a_{5}^{2}\!-36a_{0}a_{1}^{3}a_{4}^{3}a_{5}+1020a_{0}a_{1}^{2}a_{2}^{2}a_{4}a_{5}^{2}\!
+560a_{0}a_{1}^{2}a_{2}a_{3}^{2}a_{5}^{2}\!-746a_{0}a_{1}^{2}a_{2}a_{3}a_{4}^{2}a_{5}
\\
\phantom{D=}{}
+144a_{0}a_{1}^{2}a_{2}a_{4}^{4} +24a_{0}a_{1}^{2}a_{3}^{3}a_{4}a_{5}-6a_{0}a_{1}^{2}a_{3}^{2}a_{4}^{3}
-630a_{0}a_{1}a_{2}^{3}a_{3}a_{5}^{2}+24a_{0}a_{1}a_{2}^{3}a_{4}^{2}a_{5}
 \\
\phantom{D=}{} +356a_{0}a_{1}a_{2}^{2}a_{3}^{2}a_{4}a_{5}
-80a_{0}a_{1}a_{2}^{2}a_{3}a_{4}^{3}-72a_{0}a_{1}a_{2}a_{3}^{4}a_{5}
+18a_{0}a_{1}a_{2}a_{3}^{3}a_{4}^{2}+108a_{0}a_{2}^{5}a_{5}^{2}
 \\
\phantom{D=}{}
-72a_{0}a_{2}^{4}a_{3}a_{4}a_{5}+16a_{0}a_{2}^{4}a_{4}^{3}
+16a_{0}a_{2}^{3}a_{3}^{3}a_{5}-4a_{0}a_{2}^{3}a_{3}^{2}a_{4}^{2}
+256a_{1}^{5}a_{5}^{3}-192a_{1}^{4}a_{2}a_{4}a_{5}^{2}
\\
\phantom{D=}{}
-128a_{1}^{4}a_{3}^{2}a_{5}^{2}+144a_{1}^{4}a_{3}a_{4}^{2}a_{5} -27a_{1}^{4}a_{4}^{4}+144a_{1}^{3}a_{2}^{2}a_{3}a_{5}^{2}
-6a_{1}^{3}a_{2}^{2}a_{4}^{2}a_{5}-80a_{1}^{3}a_{2}a_{3}^{2}a_{4}a_{5}
 \\
\phantom{D=}{}
+18a_{1}^{3}a_{2}a_{3}a_{4}^{3}+16a_{1}^{3}a_{3}^{4}a_{5}
-4a_{1}^{3}a_{3}^{3}a_{4}^{2}-27a_{1}^{2}a_{2}^{4}a_{5}^{2}
+18a_{1}^{2}a_{2}^{3}a_{3}a_{4}a_{5}-4a_{1}^{2}a_{2}^{3}a_{4}^{3}\\
\phantom{D=}{}
-4a_{1}^{2}a_{2}^{2}a_{3}^{3}a_{5}+a_{1}^{2}a_{2}^{2}a_{3}^{2}a_{4}^{2}.
\end{gather*}
However, to write down the general case explicitly is not easy
(almost impossible).

\medskip

\noindent
{\bfseries Problem.} {\it Calculate \eqref{eq:general-integral} for $n=4$ $($and
$n=5)$ directly.}

\medskip

The wall called Gaussian is very high and not easy to overcome,
and therefore hard work will be needed.

%After submitting this paper to {\bfseries arXiv}
Recently
the subsequent paper \cite{MS2}
by Morozov and Shakirov appeared. Our works are deeply related to
so-called non-linear algebras, so we will make some comments on
this point in a near future. As a general introduction to them
see for example \cite{DM}.

The author thanks referees and Hiroshi Oike, Ryu Sasaki for many useful
suggestions and comments.

\pdfbookmark[1]{References}{ref}
\LastPageEnding


\begin{thebibliography}{99}

\footnotesize\itemsep=0pt

\bibitem{Fu}
 Fujii K.,
Beyond Gaussian: a comment,
\href{http://arxiv.org/abs/0905.1363}{arXiv:0905.1363}.

\bibitem{MS}
Morozov A., Shakirov Sh.,
Introduction to integral discriminants,
\href{http://dx.doi.org/10.1088/1126-6708/2009/12/002}{{\it J.~High Energy Phys.}} {\bf 2009} (2009), no.~12, 002, 39~pages,
\href{http://arxiv.org/abs/0903.2595}{arXiv:0903.2595}.

\bibitem{WW}
Whittaker E.T., Watson  G.N.,
A course of modern analysis,
Cambridge University Press, Cambridge, 1996.

\bibitem{Sa}
Satake I.,
 Linear algebra,
Shokabo, Tokyo, 1989  (in Japanese).

\bibitem{Fu2} Fujii K.,
Beyond the Gaussian. II.~Some applications, in progress.

\bibitem{MS2}
Morozov A., Shakirov Sh.,
New and old results in resultant theory,
\href{http://dx.doi.org/10.1007/s11232-010-0044-0}{{\it Theoret. and Math. Phys.}} {\bf 163} (2010), 587--617,
\href{http://arxiv.org/abs/0911.5278}{arXiv:0911.5278}.

\bibitem{DM}
Dolotin V., Morozov  A.,
Introduction to non-linear algebra,
World Scientif\/ic Publishing Co. Pte. Ltd., Hackensack, NJ, 2007,
\href{http://arxiv.org/abs/hep-th/0609022}{hep-th/0609022}.

\end{thebibliography}
\end{document}